\documentstyle[11pt,paspconf,epsf]{article}

\begin{document}

\title{2.5--11.6~$\mu$ spectrophotometry and imaging of the CfA sample}
\author{J. Clavel, B. Schulz, B. Altieri, P. Barr, P. Claes, A. Heras,
K. Leech, L.Metcalfe, A. Salama}
\affil{ISO Science Operations Centre, Astrophysics Division, ESA Space Science
Dept.,P.O. Box 50727, 28080 Madrid, Spain}
\begin{abstract}
We present low resolution spectrophotometric and imaging ISO observations 
of a sample of 54 AGN's over the 2.5--11~$\mu$ range. The observations generally
support unification schemes and set new constraints on models of the molecular 
torus.
\end{abstract}
\section{Introduction}
According to the so-called ``unified model'' of Active Galactic Nuclei (AGN), 
Seyfert 1 and Seyfert 2 galaxies (hereafter Sf1 and Sf2, respectively) are 
essentially the same objects viewed at a different angle: Sf1's are observed 
close to face-on such that we have a direct view to the Broad emission 
Line Region (BLR) and to the accretion disk responsible for the strong 
UV-Optical-X-ray continuum, whereas Sf2's are seen at an inclination such that 
our view is blocked by an optically thick dusty torus which surrounds 
the disk and the BLR (e.g. Antonucci 1993). This model makes specific 
predictions. In particular, the UV photons from the disk which are absorbed by 
the grains in the torus should be re-emitted as dust thermal emission in the IR.
Several observational arguments constrain the torus inner radius to be of the 
order of $\sim$ 1 pc  in which case the gas and dust temperature should peak 
to about 700--1000~K  and give rise to a thermal emission ``bump'' with a 
maximum between 7 and 15~$\mu$ depending on the torus inclination (Pier and 
Krolik 1992). This model also predicts that the silicate 9.7~$\mu$ 
feature should appear preferentially in absorption in Sf2's and in emission 
in Sf1's. In order to test these predictions and better constrain the model, 
we initiated a programme of mid-IR observations of a large sample of AGN's.
Throughout we use ${\rm H_{0}\,=\,75\,km\,s^{-1};\,q_{0}\,=\,0}$.
\section{Observations}
A sample of 54 Active Galactic Nuclei (AGN) was observed with the PHT 
(Lemke et al. 1996) and CAM (Cesarsky et al. 1996) instruments on board the 
Infrared Space Observatory (ISO; Kessler et al. 1996). The sample is 
drawn from the CfA hard X-ray flux limited complete sample but lacks the most 
well known objects (e.g. NGC~4151) which were embargoed by ISO guaranteed time 
owners. On the other hand, the sample was enriched in bright Sf2's. We therefore
caution that our sample is by no mean ``complete'' in a statistical sense. It is
about equally divided into Sf1's (26 sources) and Sf2's (28), where we define 
Sf1's as all objects of type $\leq$ 1.5 and Sf2's those whose type is $>$ 1.5. 
The mean and $r.m.s.$ redshift are $0.036\pm0.055$ and $0.017\pm0.013$ for 
Sf1's and Sf2's, respectively. 
\begin{figure}
\plottwo{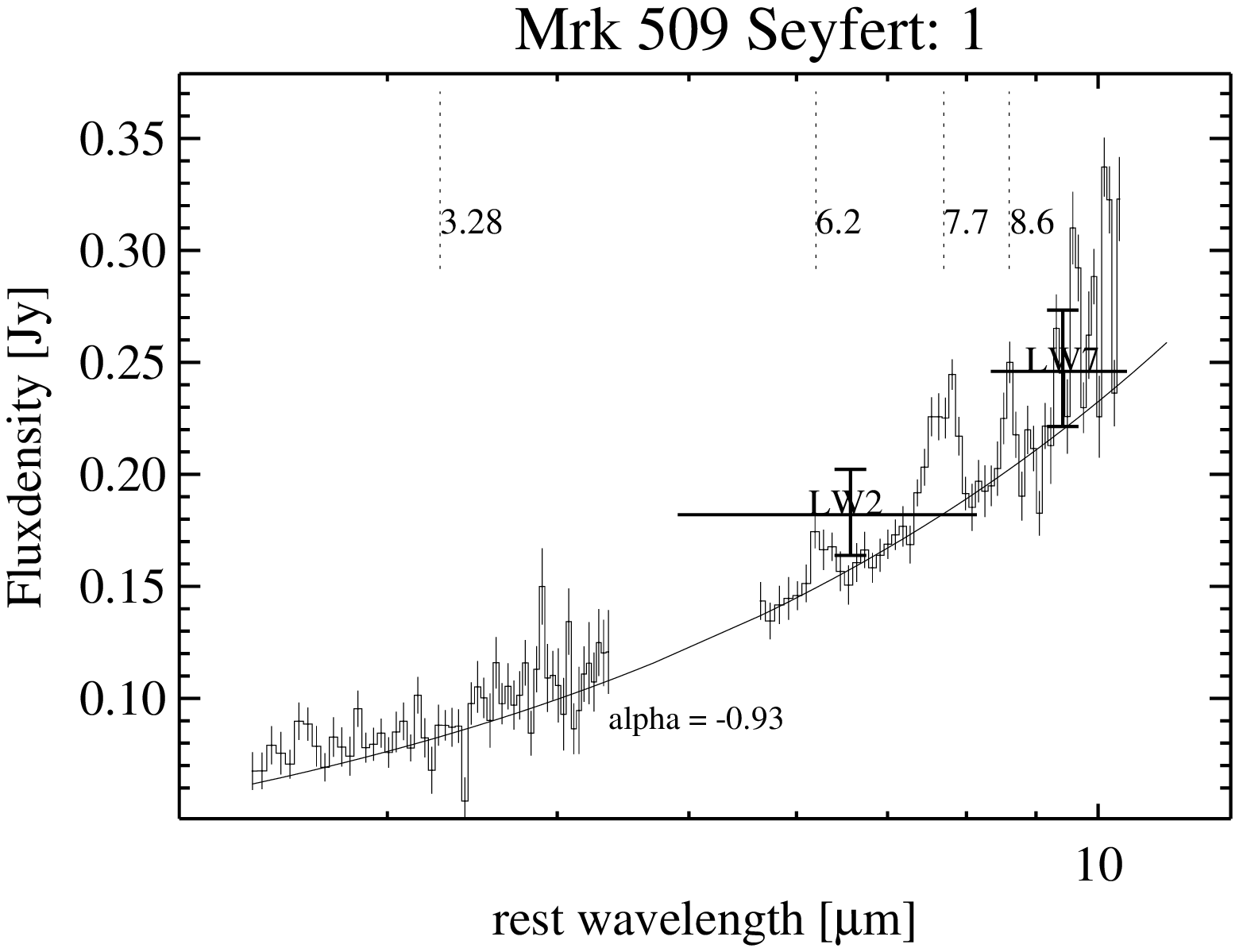}{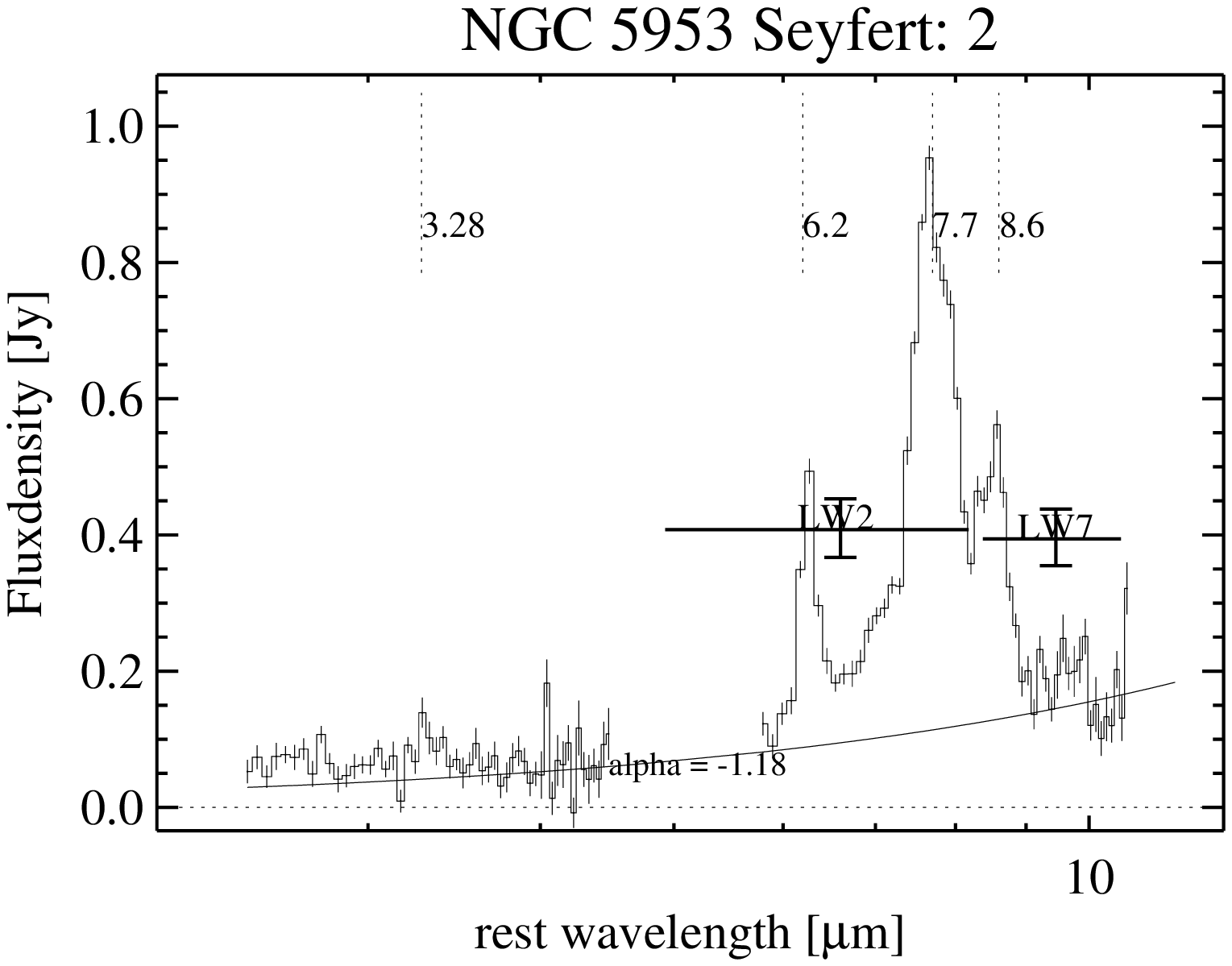}
\caption{Two representative spectra of a Sf1 (left) and a Sf2 (right) with
error bars. The two data-points marked LW2 and LW7 indicate the flux 
from the CAM images with its error and the filter wavelength range. The
best-fit power-law continuum is also shown}
\label{fig-1}
\end{figure}
For each object, the data-set consists of CAM images obtained through  
filters at 6.75 and 9.63~$\mu$ and at a magnification of 3'' per pixel, 
together with 2.5--11.6~$\mu$ spectra obtained immediately after with the 
PHT-S low resolution (${\rm 3000\,km\,s^{-1}}$) spectrograph. The images 
consists of arrays of $32\,\times\,32$ pixels (i.e. $96\,\times\,96$'') with an 
effective resolution (FWHM) of 3.8'' and 4.5'', for the 6.75 and 9.63~$\mu$
filters, respectively. The exposure times were 200~s, sufficiently
long to ensure stabilisation of the detectors. For the spectra, on-source 
measurements were alternated with sky measurements at a frequency of 1/256~Hz,
with a chopper throw of 300''. The on-source exposure times varied between 
512~s and 2048~s, depending on the source brightness. The spectrograph 
aperture is $21''\,\times\,21''$ square. In all cases, the array or 
spectrograph was centered at the nominal position of the nucleus to within
2''.
\section{Calibration and data reduction}
The CAM images were reduced and calibrated using standard procedures of the 
CAM Interactive Analysis (CIA) package. Nuclear fluxes were obtained by 
integrating all the emission in a circle of 5--6 pixels radius (15--18'').
Their accuracy, mainly limited by flat-fielding residuals, is $\pm10$~\%. 
Radial profiles were also computed and compared to that of point-like
calibration stars. For all sources but 4, we verified that the AGN 
are unresolved at the $\simeq4$'' resolution of ISOCAM.

Because PHT-S was operating close to its sensitivity limit, special reduction 
and calibration procedures had to be applied. After a change of illumination, 
the responsivity of the Si:Ga photoconductors immediately jumps to an 
intermediate level. This initial jump is followed by a characteristic slow 
transient to the final level. At the faint flux limit, this time constant is 
extremely long, and in practice only the initial step is observed in 
chopped-mode. The spectral response function for this particular mode and 
flux-level was derived directly from observations of a faint 
standard star HD~132142 whose flux ranges from 0.15 to 2.54 Jy. The 
calibration star observation was performed with the same chopper frequency 
and readout-timing as the astronomical observations. The $S/N$ of the PHT-S 
spectra was considerably enhanced by two additional measures that deviate from 
the standard reduction procedure. $i$) the long integration ramps 
(32~s each) were divided into sub-ramps of 2~sec and no de-glitching 
(removal of cosmic ray hits) was performed at ramp-level $ii$) after 
slope-fitting and de-glitching at slope-level, the maximum of the distribution 
of the slopes was determined by fitting a gaussian to the histogram.
The resulting PHT-S fluxes should be accurate to within $\pm10$~\%.
\begin{figure}
\plottwo{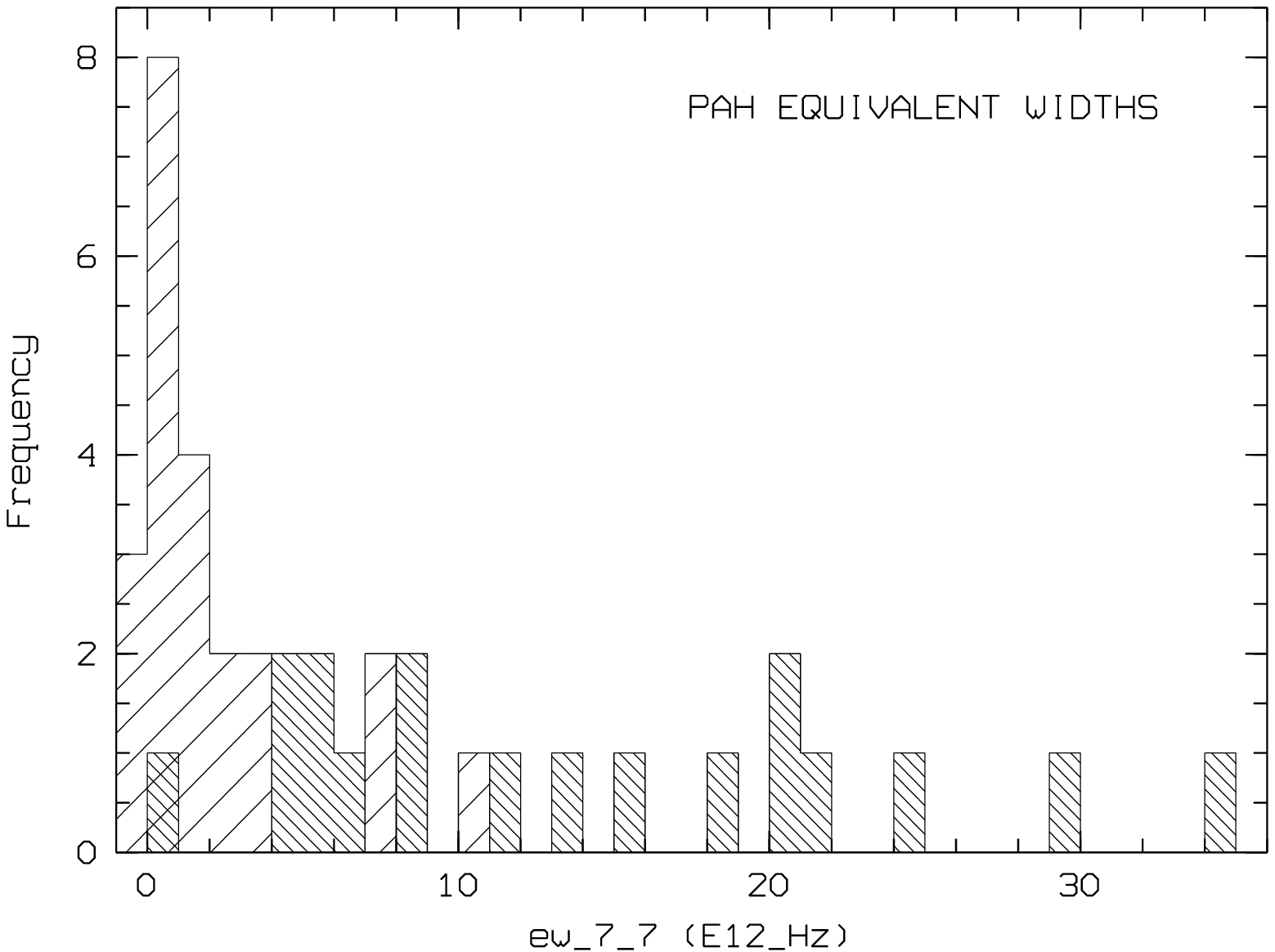}{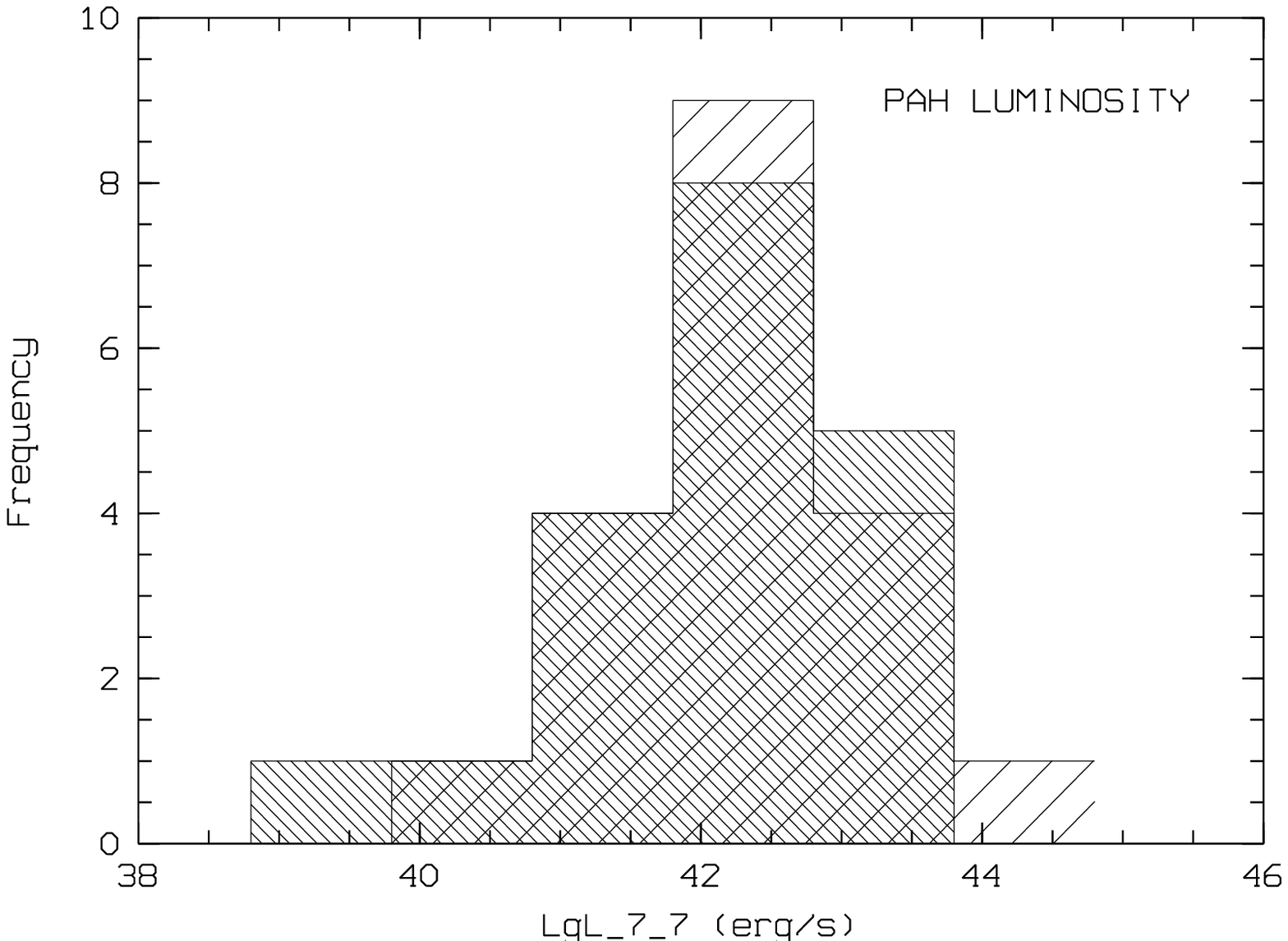}
\caption{Distribution of PAH 7.7~$\mu$ EW (left)
\& luminosities (right) for Sf1 (spaced hatching at 
+45~$\deg$) \& Sf2 (fine hatching, -45~$\deg$). 
}\label{fig-2}
\end{figure}
For all sources but 4, the agreement between the CAM and the PHT-S fluxes is 
excellent which confirms the reliability of our calibration. The 4 mismatches 
correspond to those cases where the source is spatially resolved and
contribution from the underlying galaxy is visible in the CAM image (see
above).
\section{The difference between Sf1 and Sf2}
As illustrated in fig~\ref{fig-1}, the mid-IR spectrum of a typical Sf1 
(Mrk~509, left) is markedly different from that of a Sf2 (NGC~5953, right): 
Sf1's have a strong continuum well approximated by a power-law
(${\rm F_{\nu}\,\propto\,\nu^{\alpha}}$) of average index and $r.m.s.$ 
dispersion $\langle \alpha \rangle = -0.91\pm0.25$ and weak emission features. 
By contrast, Sf2's display only a weak and ill-defined continuum 
together with very strong emission features. These features have
well defined peaks at at 6.2, 7.7 and 8.6~$\mu$, and in many Sf2's, a 
weaker feature can also be seen at 3.3~$\mu$. These 
features are ubiquitous in many galactic and extragalactic sources, in
particular HII regions and starburst galaxies (see Genzel et al 1998 for
instance). They are usually ascribed to Polycyclic Aromatic Hydrocarbon 
(PAH) bands. In many galaxies of adequate $S/N$ and redshift, the blue
side of the strong 11.3~$\mu$ PAH feature is also detected as a sharp 
rise in flux toward the long wavelength end of the PHT-S array (see e.g.
fig~3.b). 

\begin{figure}
\plottwo{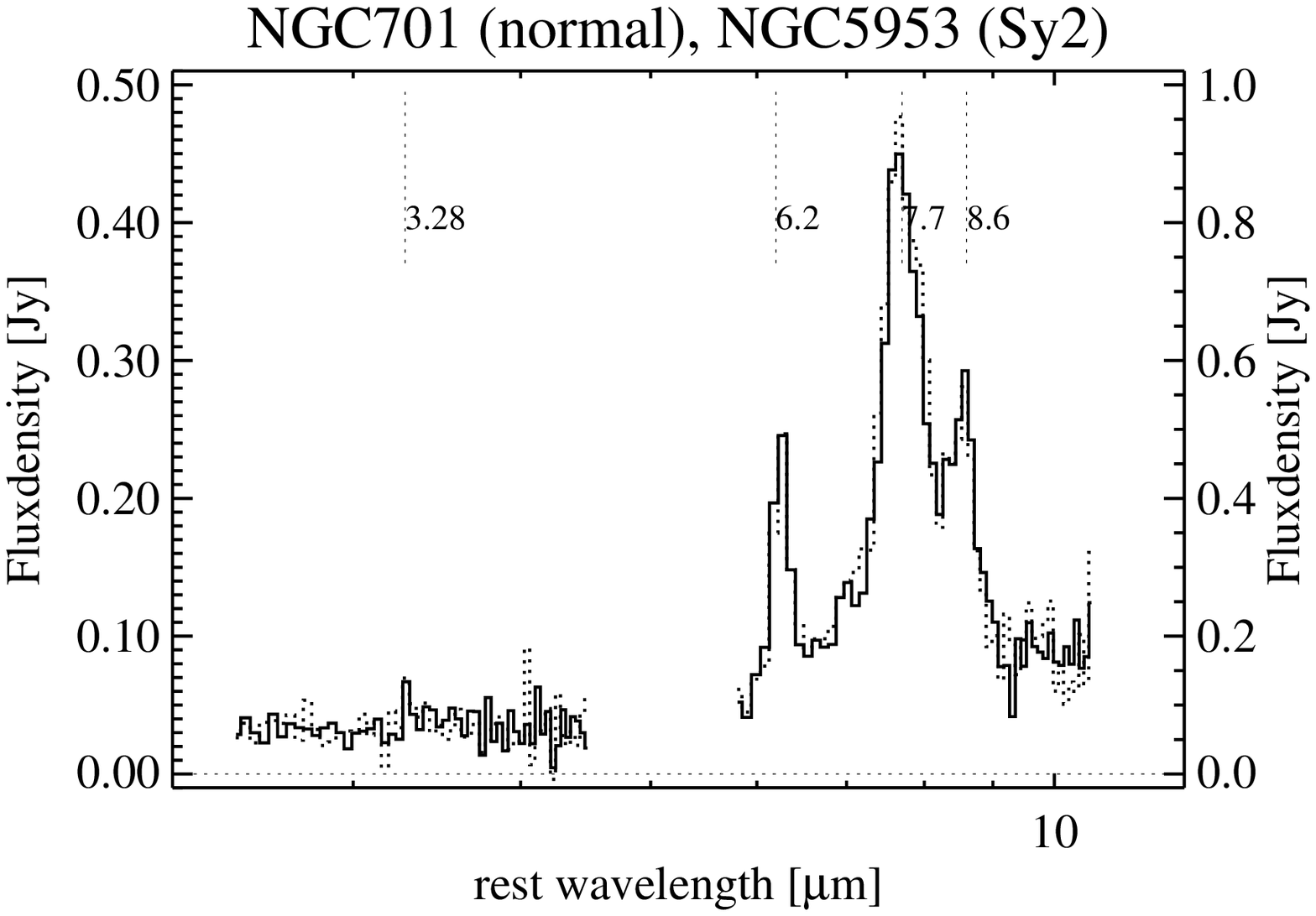}{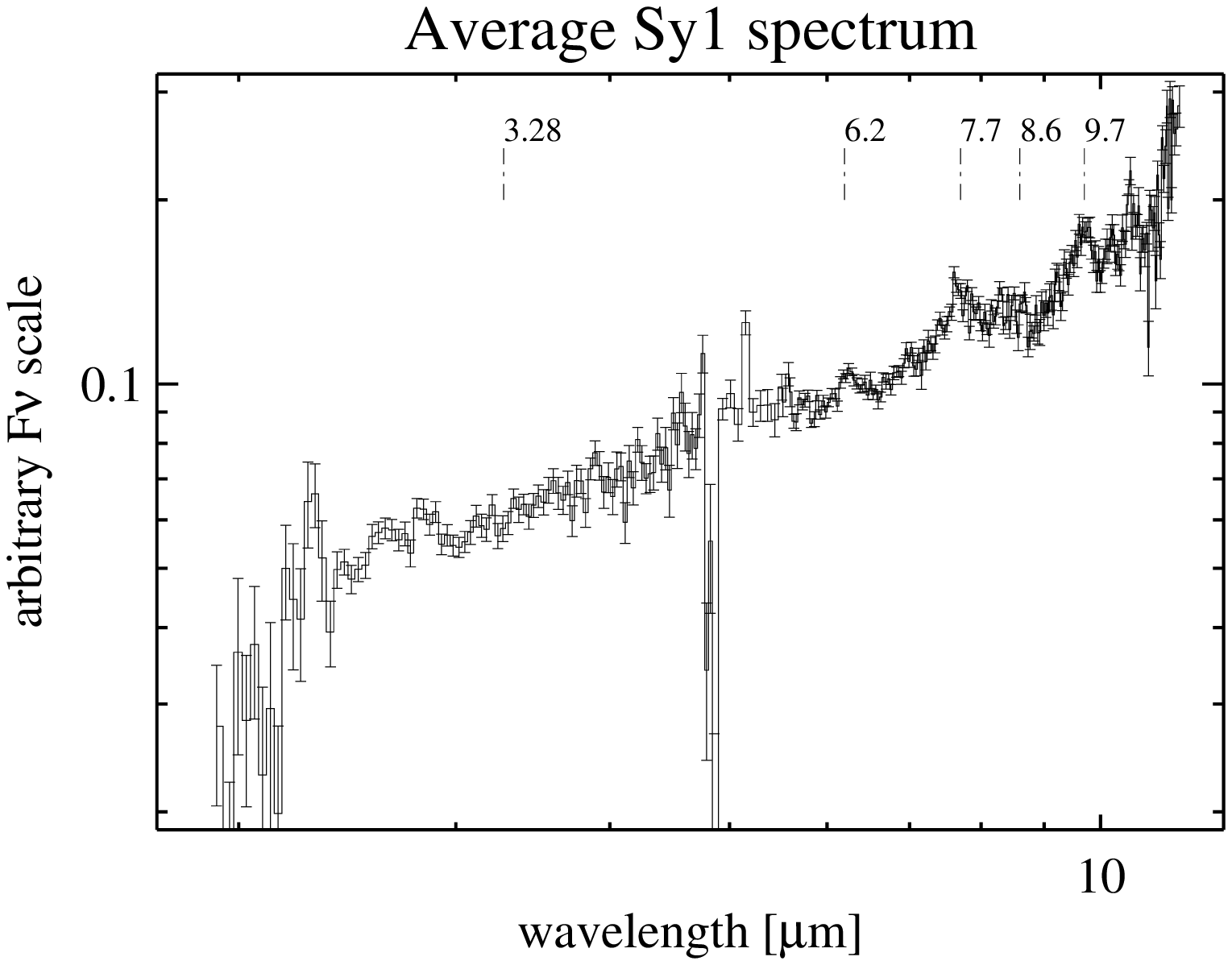}
\caption{Left: the spectra of the non-active galaxy NGC~701 
(heavy line) and of the Sf2 galaxy NGC~5953 (dotted-line).
Right: The average Sf1 spectrum of the 26 type $\leq\,1.5$ galaxies
in the sample}
\label{fig-3}
\end{figure}

The PAH fluxes and Equivalent Widths (EW) were computed by integrating all the 
emission above the best-fit power-law continuum. The resulting EW distribution
of the strongest PAH bands at 7.7~$\mu$ is shown in fig~\ref{fig-2}.a. It 
clearly illustrates that the PAH in Sf1's are systematically weaker than in 
Sf2's where EW's extend up to $\sim\,9\,\mu$. A two-tail KS test confirms that 
Sf1's and Sf2's have statistically different EW distributions at the 
$2\,10^{-7}$ confidence level. The mean ($\pm r.m.s$) PAH equivalent widths of 
Sf1's is ${\rm \langle EW_{7.7} \rangle\,=\,0.40\pm0.54\,\mu}$, nearly 7 times 
smaller than that of Sf2's, 
${\rm \langle EW_{7.7} \rangle\,=\,2.66\pm2.18\,\mu}$. As can be seen from 
fig~\ref{fig-2}.b, the distribution of the 7.7~$\mu$ PAH {\em luminosities\/} 
is however the same in Sf1's and Sf2's, at the 22~\% confidence level 
(KS test). The mean ($\pm r.m.s$) PAH luminosity of Sf1's is 
${\rm \langle \log{L_{7.7}} \rangle\,=\,42.22\pm\,0.75\,erg\,s^{-1}}$, 
not statistically different (at the 92.9~\% confidence level from a Student-t
test) from that of Sf2's, 
${\rm \langle \log{L_{7.7}} \rangle\,=\,42.19\pm0.95\,erg\,s^{-1}}$.

\section{Implication for unification schemes}

The continuum flux at a fiducial wavelength of 7~$\mu$ was read-out from
the best-fit power-law. The resulting mean ($\pm r.m.s.$) continuum 
logarithmic luminosities for Sf1's and Sf2's are 
${\rm \langle \log{\nu\,L_{\nu,7.7}} \rangle=43.74\pm0.79}$
and $42.82\pm0.78$ ${\rm erg\,s^{-1}}$, 
respectively, implying that the mid-IR continuum of Sf2's is $\sim$~8 times
less luminous, on the average, than that of Sf1's. This ratio is very close
to the Sf2/Sf1 EW ratio. Altogether, these results imply that {\em the prime
distinguishing feature of Sf2's in the mid-IR is that their continuum is 
depressed relative to that of Sf1's.}

The above result is broadly consistent with unification schemes in that the 
mid-IR continuum, which is directly visible in face-on objects (i.e. Sf1's),
is largely extinguished in edge-on objects (i.e. Sf2's). It further implies 
that the torus is largely opaque to its own mid-IR radiation. Assuming that 
face-on objects suffer zero extinction, one can use the average ratio 
$\langle R \rangle$ of PAH EW in Sf1's and Sf2's to infer the mean 
7.7~$\mu$ extinction along the direction perpendicular to the torus symmetry 
axis (i.e. radially). The average ratio is $\langle R \rangle\,=\,6.7\pm5.5$, 
where the error quoted reflects the r.m.s. dispersion of Sf2 EW's (we 
have neglected the scatter in Sf1's EW's). This implies that the Sf2 
continuum suffers on the average from $2.1\pm0.9$ magnitudes of extinction 
at 7.7~$\mu$, i.e. ${\rm A_{v}\,=\,103\pm44}$ magnitudes 
(Rieke and Lebofsky 1985). For a normal gas to dust ratio, this corresponds to
an average X-ray absorbing column along one torus radius, 
${\rm N_{H}\,=\,2.3\pm1.0\,10^{23}\,cm^{-2}}$ (Gorenstein 1975). The latter 
is in good agreement with the mean absorbing column of 8 Sf2's (excluding the
lower limits on NGC~1068 and NGC~7582) as measured directly from X-ray data, 
${\rm N_{H}\,=\,1.1^{+2.9}_{-0.8}\,10^{23}\,cm^{-2}}$ (Mulchaey et al. 1992). 

It also suggests that PAH emission is isotropic and arises from outside the 
torus, either in the Narrow Line Region or in the ISM of the bulge. 
To check the origin of the PAH features in Sf2's, we have observed the 
nucleus of the ``normal'' (i.e. non active)
SB galaxy NGC~701. Its PHT-S spectrum is plotted in fig~\ref{fig-3}.a together 
with that of the Sf2 nucleus NGC~5953. The two spectra are virtually 
indistinguishable. As a matter of fact, the EW of the PAH 7.7~$\mu$ feature 
in NGC~701, 4.2$\pm0.4\,\mu$ and its luminosity
${\rm \log{L_{7.7}}\,=\,42.012\pm0.004\,erg\,s^{-1}}$ are in the middle of the 
range of Sf2's. We therefore conclude that the {\em PAH emission is 
unrelated to the active nucleus and arise from the ISM in the bulge.} 

\begin{figure}
\plotone{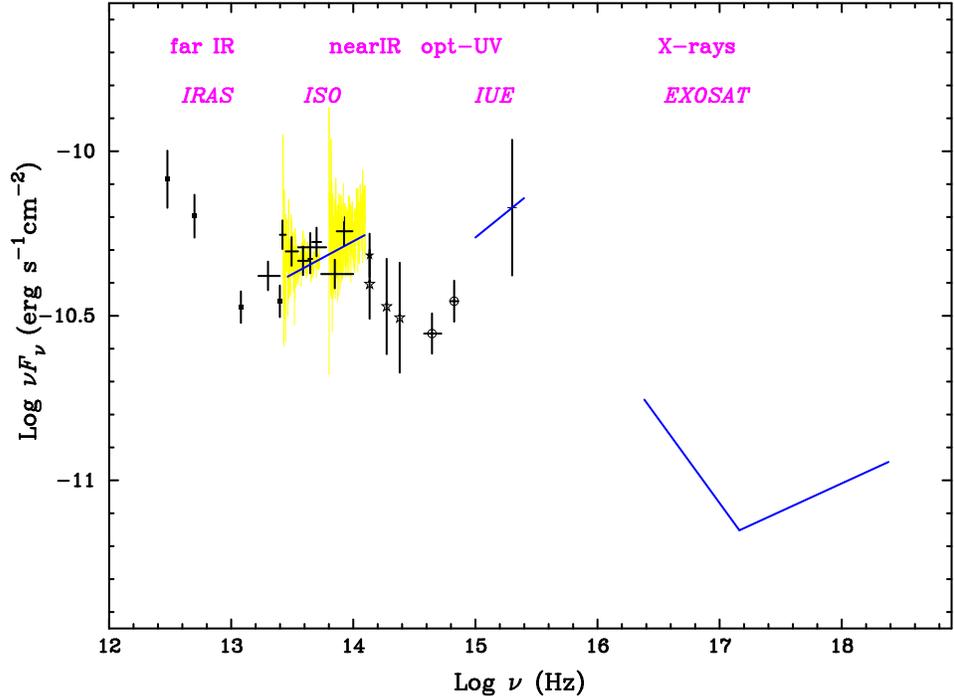}
\caption{The Mrk~279 SED from 100~$\mu$ to 10~keV.}
\label{fig-4}
\end{figure}

As can be seen from fig~\ref{fig-3}.a, the continuum of NGC~5953 is also 
indistinguishable from that of NGC~701. This further suggests that in Sf2's 
with PAH EW's in excess of $\simeq4\,\mu$, the torus emission is completely 
suppressed, at least up to 11~$\mu$, and the faint residual mid-IR continuum 
arises entirely from outside the active nucleus, i.e. from stars at 
wavelengths shorter than $\sim\,3\,\mu$ and from small ISM dust particles at 
longer wavelengths. In such objects, the PAH EW only provides a lower limit 
to ${\rm A_{v}}$ and ${\rm N_{H}}$. Hence, our previous estimate of the 
average extinction in Sf2's is probably underestimated. From the 28 Sf2's, 
8 have ${\rm EW_{7.7}\,\geq\,4\,\mu}$. This suggests that about
30~\% of Sf2 suffer from extinction in excess of 125 visual magnitudes and 
have hydrogen columns larger than ${\rm 3\,10^{23}\,cm^{-2}}$. These extreme 
Sf2's probably correspond to NGC~1068-type objects where the absorbing
column is so large as to block all hard X-rays.

\section{The Silicate 9.7~$\mu$ features and further constraints on the torus}

Fig.~\ref{fig-3}.b shows the weighted mean Sf1 spectrum obtained by averaging 
the rest wavelength spectra of all 26 type $\leq\,1.5$ galaxies. The Silicate 
9.7~$\mu$ feature is present {\em in emission\/} with an equivalent width 
${\rm \langle EW_{9.7} \rangle\,=\,0.236\pm0.008\,\mu}$. This immediately
{\em rules out models with very large torus optical depths.\/} In the model of
Pier and Krolik (1992) for instance, the strength of the Silicate feature
is calculated as a function of inclination $i$ and of the vertical and radial 
Thomson optical depth $\tau_{z}$ and $\tau_{r}$, respectively. Reading from 
their figure 8, models with $\tau_{z}\,\geq\,1$ and/or $\tau_{r}\,\geq\,1$
are ruled-out as they would predict the Silicate feature in absorption.
For an average Sf1 inclination $\cos{i}\,=\,0.8$, the best fit to 
${\rm \langle EW_{9.7} \rangle\,=\,0.236\pm0.008\,\mu}$ suggests
$\tau_{r}\,\simeq\,1$ and $0.1 \leq \tau_{z}\,\leq\,1$. A unit Thomson 
optical depth corresponds to a column density 
${\rm N_{H}\,=\,10^{24}\,cm^{-2}}$. While these figures are somewhat uncertain
and model dependent, it is reassuring that the inferred range in column 
densities agrees with our previous estimate based on the Sf2 extinction.

We are not able to state whether the Silicate feature is present in Sf2's.
As a matter of fact, the PAH bands at 7.7, 8.6 and 11.3~$\mu$ are so strong 
that placing the continuum at 9.7~$\mu$ becomes a rather subjective
decision. In the absence of longer wavelengths data, the best we can
do is set a provisional upper limit of 1~$\mu$ to the Silicate EW in Sf2's, 
whether in absorption or in emission. 

\section{Energy distribution of Seyfert 1's}

The SED of Mrk~279 is shown in fig~\ref{fig-4}. This Sf1 galaxy was also the 
subject of an ISO monitoring campaign the result of which will be presented 
elsewhere (Santos-Ll\'eo et al., 1998). Its spectral index 
$\alpha = -0.80\pm0.05$ and PAH 7.7~$\mu$ EW ($0.076\pm0.024\,\mu$) are fairly
typical of Sf1's. The Mrk~279 SED was assembled from published data as explained
in Santos-Ll\'eo et al. The average IUE flux is plotted with its mean 
spectral index and range of variability. The PHT-S spectrum is shown together
with the best fit power-law continuum and the ISOCAM photometric data-points.
Beside the usual ``big blue bump'' and the far-IR ($\lambda \geq 25\,\mu$) 
rise due to cool dust emission from the large scale ISM, the Mrk~279 SED 
shows a third {\em excess in the range 1.25--15~$\mu$.\/} As noted by 
Santos-Ll\'eo et al., the spectral shape of this ``mid-IR bump'' is strikingly 
similar to the theoretical emission spectrum of a face-on torus as computed by 
Pier and Krolik (1992). A detailed comparison with the model is beyond the 
scope of this paper, but we take this match as a further confirmation that 
unification schemes are correct, at least to a first order approximation, 
and that a molecular torus exists in at least some AGN's. On the other hand,
it is not clear that all Sf2's are obscured by a thick torus. Indeed, those
Sf2's with small PAH EW -- similar to those of Sf1's (see fig-\ref{fig-2}.a) --
could well be genuine Sf1's where our line of sight intercepts a foreground
dust screen in the general ISM of the host galaxy.

\begin{question}{Dr.\ Deborah Dultzin-Hacyan} Have you compared the properties
of Sf2's with starburst galaxies ? 
\end{question} 
\begin{answer}{Jean Clavel}
Unfortunately there were no starburst galaxies in our sample. However,  Genzel
et al (1998) have observed a sample of 12 starburst galaxies  with PHT-S. They
define a``PAH strength'' as the ratio of the flux at the  peak of the 7.7~$\mu$
feature to the underlying continuum flux. The mean ($\pm$ r.m.s.) strength for
their starburst's sample is $3.8\pm2.6$, almost identical to that of our Sf2
sample, $3.6\pm2.8$.
\end{answer}

\begin{question}{Dr.\ Bev Wills}
Are your results consistent with the findings of Roche and Aitken who 
concluded that the 10~$\mu$ (20~$\mu$ ?) features were present in Sf2's,
but absent in Sf1's and that this may indicate destruction of the carrier
by exposure to the AGN continuum ?
\end{question}
\begin{answer}{Jean Clavel}
Roche et al. (1991) observed a sample of 60 IR bright galaxies -- 
starburst's and AGN's -- in the 8--13~$\mu$ range. A direct
comparison with their results is difficult given the small overlap in
wavelengths between the two data-set. Their $S/N$ and resolution was also lower
than ours. One of Roche et al's conclusion was that, in AGN's, the PAH features 
``are seen only rarely, but the silicate absorption band is present in about
half of the sample and is more prevalent in the lower luminosity objects''.
This clearly conflicts with our results which show that PAH features are 
present in all AGN's but are definitely weak in Sf1's, probably too weak
to be detected from the ground. The other problem with the Roche et al.
data was their restricted wavelength range which did not allow them to 
properly define a baseline continuum. This is probably the reason why they
wrongly identified a gap between the strong PAH 7.7/8.6~$\mu$ emission blend 
and the 11.3~$\mu$ PAH feature as due to silicate 9.7~$\mu$ absorption.
\end{answer}


\begin{references}

\reference Antonucci, R. 1993, \araa, 31, 473
\reference Cesarsky, C.J., et al. 1996, \aap, 315, L32
\reference Genzel, R. et al. 1998, \apj, in press (MPE Preprint 426)
\reference Gorenstein, P. 1975, \apj, 198, 95
\reference Kessler, M.F., Steinz, J.A., Anderegg, M.E., et al. 1996, \aap, 315, L27
\reference Lemke, D., Klaas, U., et al. 1996, \aap, 315, L64
\reference Mulchaey, J.S., Mushotzky, R.F. \& Weaver, K.A. 1992, \apj, 390, L69
\reference Pier, E.A. \& Krolik, J.H. 1992, \apj, 401, 99
\reference Rieke, G.H. \& Lebofski, M.J. 1985, \apj, 288, 618
\reference Roche, P.F. et al. 1991, \mnras, 248, 606. 
\reference Santos-Ll\'eo, M. et al. 1998, in preparation

\end{references}
\end{document}